\documentclass[12pt]{article}
\usepackage{epsfig}
\usepackage{amssymb,amsmath}

\newcommand{\bea}{\begin{eqnarray}}
\newcommand{\eea}{\end{eqnarray}}

 \makeatletter
\def\alt{\mathrel{\mathpalette\gl@align<}}
\def\agt{\mathrel{\mathpalette\gl@align>}}
\def\gl@align#1#2{\lower.6ex\vbox{\baselineskip\z@skip\lineskip\z@
\ialign{$\m@th#1\hfil##\hfil$\crcr#2\crcr\sim\crcr}}} \makeatother

\begin{document}
%
\vspace*{1.0cm}

\begin{center}
\baselineskip 20pt 
{\Large\bf 
Simple brane-world inflationary models \\ in light of BICEP2
}
\vspace{1cm}

{\large 
Nobuchika Okada$^{~a}$  and  Satomi Okada
}
\vspace{.5cm}

{\baselineskip 20pt \it
$^a$Department of Physics and Astronomy, University of Alabama, Tuscaloosa, AL35487, USA
} 

\vspace{.5cm}

\vspace{1.5cm} {\bf Abstract}
\end{center}

Motivated by the recent CMB $B$-mode observation announced by the BICEP2 collaboration, 
  we study simple inflationary models in the Randall-Sundrum brane-world cosmology. 
Brane-world cosmological effects alter the inflationary predictions of the spectral index ($n_s$) and 
 the tensor-to-scalar ratio ($r$) from those obtained in the standard cosmology.
In particular, the tensor-to-scalar ratio is enhanced in the presence of the 5th dimension,  
 and simple inflationary models which predict small $r$ values in the standard cosmology 
 can yield $r$ values being compatible with the BICEP2 result, $r=0.2^{+0.07}_{-0.05}$. 
Confirmation of the BICEP2 result and more precise measurements of $n_s$ and $r$ in the near future 
   allow us to constrain the 5-dimensional Planck mass ($M_5$) of the brane-world scenario. 
We also discuss the post inflationary scenario, namely, reheating of the universe through inflaton decay 
  to the Standard Model particles. 
When we require the renormalizability of the inflationary models, the inflaton only couples with the Higgs doublet 
 among the Standard Model fields, through which reheating process occurs. 
If $M_5$ is as low as $10^6$ GeV, the mass of inflaton becomes of ${\cal O}$(100 GeV), 
   providing the inflationary predictions being consistent with the BICEP2 and Planck results. 
Such a light inflaton has some impacts on phenomenology at the electroweak scale 
  through the coupling with the Higgs doublet. 
\thispagestyle{empty}

\newpage

\addtocounter{page}{-1}

\baselineskip 18pt

\section{Introduction} 
Inflationary universe is the standard paradigm in the modern cosmology~\cite{inflation1, inflation2, chaotic_inflation, inflation4} 
 which provides not only solutions to various problems in the standard big-bang cosmology, 
 such as the flatness and horizon problems, but also the primordial density fluctuations 
 as seeds of the large scale structure observed in the present universe.  
Various inflation models have been proposed with typical inflationary predictions for the primordial perturbations. 
Recent cosmological observations by, in particular, the Wilkinson Microwave Anisotropy Probe (WMAP)~\cite{WMAP9} 
 and the Planck satellite~\cite{Planck2013} experiments have measured the cosmological parameters precisely and 
 provided constraints on the inflationary predictions for the spectral index ($n_s$), 
 the tensor-to-scalar ratio ($r$), the running of the spectral index ($\alpha=d n_s/d \ln k$), 
 and non-Gaussianity of the primordial perturbations. 
Future cosmological observations are expected to be more precise towards discriminating inflationary models.

Recently the Background Imaging of Cosmic Extragalactic Polarization (BICEP2) collaboration 
 has reported the observation of CMB $B$-mode polarization~\cite{BICEP2}, 
 which is interpreted as the primordial gravity waves with $r=0.20^{+0.07}_{-0.05}$ (68\% confidence level)
 generated by inflation.  
However, future observations of the $B$-mode polarization with different frequencies are crucial to verify the BICEP2 result,  
 since uncertainty of dust polarization could dominate the excess observed by the BICEP2 experiment~\cite{dust_uncertainty}. 
If confirmed, the BICEP2 result is the direct evidence of the cosmological inflation in the early universe and 
  provide us with great progress in understanding inflationary universe. 
The observed value of the tensor-to-scalar ratio favors the chaotic inflation model~\cite{chaotic_inflation}. 
In the light of the BICEP2 result of the large tensor-to-scalar ratio of ${\cal O}(0.1)$,
   various inflation models and their predictions have been reexamined/updated. 
In this paper, we study simple inflationary models in the context of the brane-world cosmology 
  (see, for example, Ref.~\cite{OSS2014} for an update of the inflationary predictions of simple models 
   in the standard cosmology).

The brane-world cosmology is based on a 5-dimensional model first proposed 
 by Randall and Sundrum (RS) \cite{RS2}, the so-called RS II model, 
 where the standard model particles are confined on a "3-brane"  at a boundary embedded 
 in 5-dimensional anti-de Sitter (AdS) space-time. 
Although the 5th dimensional coordinate tends to infinity,  
 the physical volume of the extra-dimension is finite because of the AdS space-time geometry. 
The 4-dimensional massless graviton is localizing around the brane on which the Standard Model 
 particles reside, while the massive Kaluza-Klein gravitons are delocalized toward infinity, 
 and as a result, the 4-dimensional Einstein-Hilbert action is reproduced at low energies. 
 Cosmology in the context of the RS II setup has been intensively studied~\cite{braneworld} 
 since the finding of a cosmological solution in the RS II setup~\cite{RS2solution}. 
Interestingly, the Friedmann equation in the RS cosmology leads to a non-standard 
 expansion law of our 4-dimensional universe at high energies, 
 while reproducing the standard cosmological law at low energies. 
This non-standard evolution of the early universe causes modifications of  
 a variety of phenomena in particle cosmology, such as the dark matter relic abundance~\cite{DM_BC} 
 (see also \cite{DM_GB} for the modification of dark matter physics in a more general brane-world cosmology, 
  the Gauss-Bonnet brane-world cosmology~\cite{GB}), 
  baryogensis via leptogenesis~\cite{LG_BC}, and gravitino productions in the early universe~\cite{gravitino_BC}.

The modified Friedmann equation also affects inflationary scenario. 
A chaotic inflation with a quadratic inflaton potential has been examined in \cite{inflation_BC} 
  and it has been shown that the inflationary predictions are modified from those in the 4-dimensional standard cosmology. 
Remarkably, the power spectrum of tensor fluctuation is found to be enhanced in the presence of the 5-dimensional bulk ~\cite{PT_BC}. 
Taking this brane-cosmological effect into account, the textbook chaotic inflation models with the quadratic 
  and quartic potentials have been analyzed in \cite{inflation_models_BC}.

In the next section, we first update the inflationary predictions of the textbook inflationary models 
 for various values of the 5-dimensional Planck mass ($M_5$) in the light of the BICEP2 result.  
Next we analyze the Higgs potential and the Coleman-Weinberg potential modes with various values 
  of the inflaton vacuum expectation value (VEV) and $M_5$. 
We will show that the brane-world cosmological effect dramatically alters the inflationary predictions 
  from those in the standard cosmology. 
In particular, the tensor-to-scalar ratio which is predicted to be small in the standard cosmology can be enhanced 
  so as to be compatible with the BICEP2 result when $M_5$ is much smaller than the 4-dimensional Planck mass. 
For such small $M_5$ values, we will find a relation between the inflaton mass and the 5-dimensional Planck mass 
  to be $m \simeq 10^{-4} M_5$, while the inflationary predictions are consistent with the cosmological observations. 
Thus, when $M_5$ is as low as $10^6$ GeV, a realistic inflationary scenario can be obtained 
  even for the inflaton mass of ${\cal O}$(100 GeV).

We will also discuss reheating after inflation through the inflaton decay to the Standard Model particles. 
At the renormalizable level, the inflaton only couples to the Higgs doublet among the Standard Model fields, 
  so that this coupling plays a crucial role in reheating process. 
When an inflaton mass is much higher than the electroweak scale as usual in the standard inflationary models, 
  the inflaton decays to a pair of Higgs doublets to thermalize the universe.   
When an inflaton is lighter than the Standard Model Higgs boson, the inflaton mainly decays to a pair of bottom quarks 
  through a mass mixing with the Higgs boson. 
Such a light inflaton has an impact on Higgs boson phenomenology, since the Higgs boson can decay to a pair of inflatons. 
Precision measurement of the Higgs boson properties at future collider experiments, such as the International $e^+ e^-$ 
  Linear Collider (ILC), may reveal the existence of an extra scalar field produced through the Higgs boson decay.  
In the context of the brane-world cosmology, this scalar field can play the role of inflaton, 
  providing the inflationary predictions being consistent with the current cosmological observations. 
Another possibility we will discuss with the light inflaton is a unified picture of the inflaton and the dark matter particle. 
Introducing a $Z_2$ parity to ensure the stability of the inflaton, the light inflaton can play the role of 
  the so-called Higgs portal dark matter. 
Since the inflaton cannot decay in this scenario, we consider preheating to transmit the inflaton energy density 
  into the radiation by explosive production of the Higgs doublets via parametric resonance 
  during the oscillation of the inflaton around its potential minimum. 
Then, the universe is thermalized by the decay products of the Higgs doublet.

\section{Simple inflationary models in the brane-world cosmology}
In the RS II brane-world cosmology, the Friedmann equation for a spatially flat universe 
  is found to be~\cite{RS2solution}
\begin{equation}
H^2 = \frac{\rho}{3 M_P} \left(1+\frac{\rho}{\rho_0} \right) + \frac{C}{a^4},
\label{BraneFriedmannEq}
\end{equation}
where $M_P=2.435 \times 10^{18}$ GeV is the reduced Planck mass, 
\begin{eqnarray}
\rho_0 = 12 \frac{M_5^6}{M_P^2},
\label{rho_0}
\end{eqnarray}
with $M_5$ being the 5-dimensional Planck mass, 
 a constant $C$ is referred to as the ``dark radiation,'' 
 and we have omitted the 4-dimensional cosmological constant. 
Note that the Friedmann equation in the standard cosmology is reproduced for  $\rho/\rho_0 \ll 1$ and $\rho/(3 M_P^2) \gg C/a^4$. 
There are phenomenological, model-independent constraints for these new parameters 
 from Big Bang Nucleosynthesis (BBN), which provides successful explanations for synthesizing light nuclei in the early universe. 
In order not to ruin the success, the expansion law of the universe must obey the standard cosmological one 
 at the BBN era with a temperature of the universe $T_{\rm BBN} \simeq 1$ MeV.  
Since the constraint on the dark radiation is very severe~\cite{dark_radiation}, we simply set $C=0$ in the following analysis.
We estimate a lower bound on $\rho_0$ by $\rho_0^{1/4} > T_{\rm BBN}$ and find $M_5 > 8.8$ TeV.\footnote{ 
A more sever constraint is obtained from the precision measurements of the gravitational law in sub-millimeter range. 
Through the vanishing cosmological constant condition, we find $\rho_0^{1/4} > 1.3$ TeV, 
 equivalently, $M_5 > 1.1 \times 10^8$ GeV as discussed in the original paper by Randall and Sundrum~\cite{RS2}.
However we note that this constraint is model-dependent in general and can be moderated 
 when, for example, a scalar field is introduced in the 5-dimensional bulk (see, for example, \cite{maeda_wands}).
Hence, in this paper we impose the model-independent BBN bound, $M_5 > 8.8$ TeV.  
}
The energy density of the universe is high enough to satisfy $\rho/\rho_0 \gtrsim1$, 
 the expansion law becomes non-standard, and this brane-world cosmological effect alters 
 results obtained in the context of the standard cosmology.

Let us now consider inflationary scenario in the brane-world cosmology with the modified Friedmann equation. 
In slow-roll inflation, the Hubble parameter is approximately given by (from now on, we use the Planck unit, $M_P=1$)
\begin{eqnarray}
  H^2 = \frac{V}{3} \left( 1+ \frac{V}{\rho_0}\right), 
\label{H_BC}
\end{eqnarray}
  where $V(\phi)$ is a potential of the inflaton $\phi$.   
Since the inflaton is confined on the brane, the power spectrum of scalar perturbation obeys  
  the same formula as in the standard cosmology, except for the modification of the Hubble parameter~\cite{inflation_BC}, 
\begin{eqnarray}
   {\cal P}_{\cal S} =\frac{9}{4 \pi^2} \frac{H^6}{(V')^2}, 
\end{eqnarray}
 where the prime denotes the derivative with respect to the inflaton field $\phi$. 
The Planck satellite experiment~\cite{Planck2013} constrains the power spectrum as 
$ {\cal P}_{\cal S}(k_0) = 2.215 \times 10^{-9}$ for the pivot scale chosen at $k_0=0.05$ Mpc$^{-1}$.  
The spectral index is given by
\begin{eqnarray}
 n_s -1 = \frac{d \ln {\cal P}_{\cal S}}{d\ln k} =-6 \epsilon + 2 \eta 
\end{eqnarray}
 with the slow-roll parameters defined as  
\begin{eqnarray}
\epsilon = \frac{V^\prime}{6 H^2} \left( \ln H^2 \right)^\prime, \; \; \eta =\frac{V^{\prime \prime}}{3 H^2} .
\end{eqnarray}
The running of the spectral index, $\alpha=dn_s/d\ln k$,  is given by 
\begin{eqnarray}
  \alpha=\frac{dn_s}{d\ln k} =\frac{V^\prime}{3 H^2}  \left( 6 \epsilon^\prime - 2 \eta^\prime   \right).  
\end{eqnarray}
On the other hand, in the presence of the extra dimension where graviton resides, 
  the power spectrum of tensor perturbation is modified to be~\cite{PT_BC} 
\begin{eqnarray}
 {\cal P}_{\cal T} =8  \left(  \frac{H}{2 \pi} \right)^2 F(x_0)^2, 
 \label{PT}
 \label{P_T}
\end{eqnarray}
where  $x_0 = 2 \sqrt{3 H^2/\rho_0}$, and 
\begin{eqnarray}
 F(x)= \left( \sqrt{1+x^2} - x^2 \ln\left[ \frac{1}{x}+\sqrt{1+\frac{1}{x^2}} \right]   \right)^{-1/2}. 
\end{eqnarray} 
For $x_0 \ll 1$ (or $V/\rho_0 \ll1$), $F(x_0) \simeq 1$, and Eq.~(\ref{P_T}) reduces to 
 the formula in the standard cosmology. 
For $x_0 \gg 1$ (or $V/\rho_0 \gg1$), $F(x_0) \simeq \sqrt{3 x_0/2} \simeq \sqrt{3 V/\rho_0} \gg 1$. 
The tensor-to-scalar ratio is defined as $r =  {\cal P}_{\cal T}/ {\cal P}_{\cal S}$.

The e-folding number is given by 
\begin{eqnarray}
N_0 = \int_{\phi_e}^{\phi_0} d\phi \; \frac{3 H^2}{V^\prime} = \int_{\phi_e}^{\phi_0} d\phi \; \frac{V}{V^\prime} \left(1+\frac{V}{\rho_0} \right), 
\end{eqnarray}
where $\phi_0$ is the inflaton VEV at horizon exit of the scale corresponding to $k_0$, 
  and $\phi_e$ is the inflaton VEV at the end of inflation, which is defined by ${\rm max}[\epsilon(\phi_e), | \eta(\phi_e)| ]=1$.
In the standard cosmology, we usually consider $N_0=50-60$ in order to solve the horizon problem. 
Since the expansion rate in the brane-world cosmology is larger than the standard cosmology case, 
  we may expect a larger value of the e-folding number. 
For the model-independent lower bound,  $\rho_0^{1/4} > 1$ MeV, the upper bound $N_0 <75$ 
 was found in \cite{N_BC}. 
In what follows, we consider $N_0=50$, $60$, and $70$, as reference values.

\subsection{Textbook inflationary models}
We first analyze the textbook chaotic inflation model with a quadratic potential~\cite{chaotic_inflation},  
\begin{eqnarray}
V =\frac{1}{2} m^2 \phi^2. 
\end{eqnarray}
In the standard cosmology, simple calculations lead to the following inflationary predictions:
\begin{eqnarray}
n_s=1-\frac{4}{2 N_0+1},  \; \; 
r=\frac{16}{2 N_0+1}, \; \; 
\alpha= - \frac{8}{(2 N_0+1)^2}. 
\end{eqnarray}
The inflaton mass is determined so as to satisfy the power spectrum measured by the Planck satellite experiment, 
 ${\cal P}_{\cal S}(k_0)=2.215 \times 10^{-9}$:
\begin{eqnarray}
 m [{\rm GeV}]= 1.46 \times 10^{13}  \left(  \frac{60.5}{N_0+1/2}\right). 
\end{eqnarray}

\begin{figure}[htbp]
\begin{center}
\includegraphics[width=0.45\textwidth,angle=0,scale=1.04]{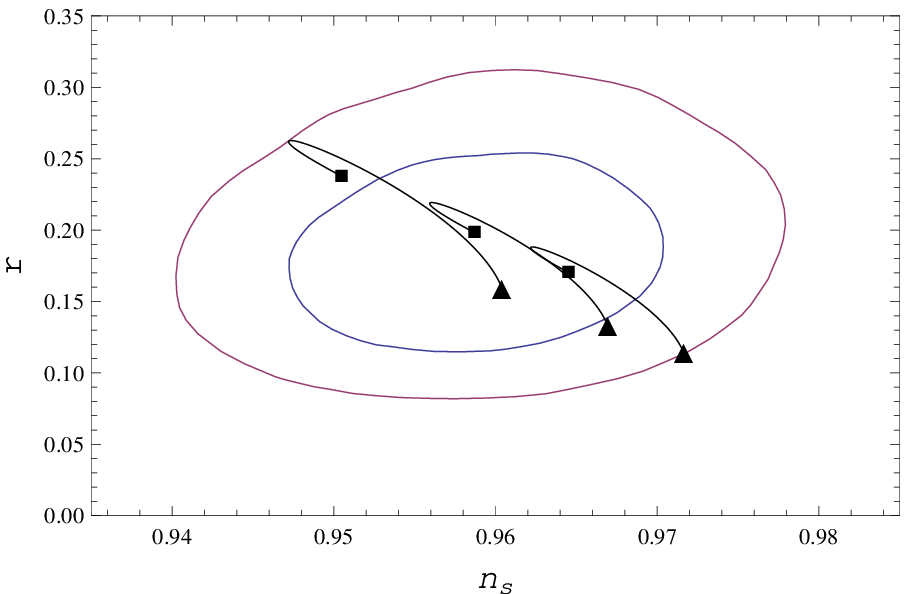} \hspace{0.5cm}
\includegraphics[width=0.45\textwidth,angle=0,scale=1.08]{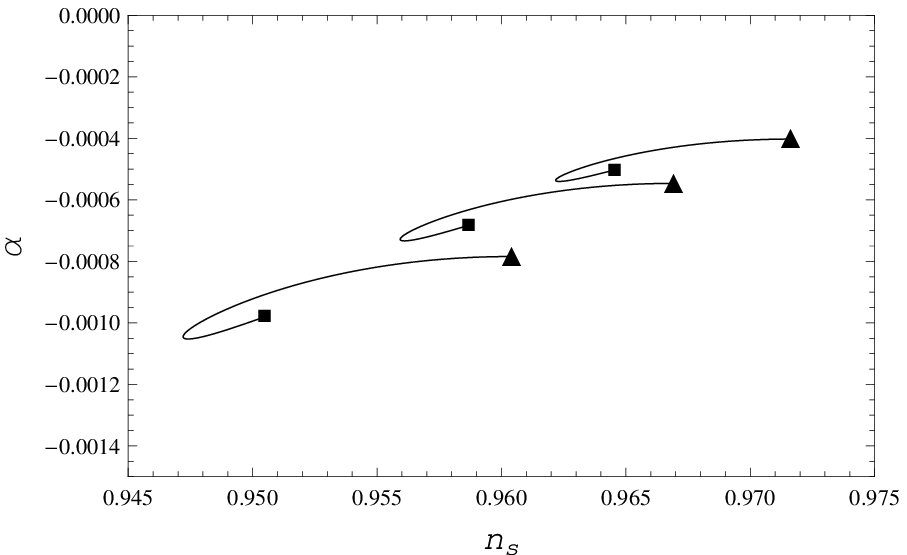}
\end{center}
\caption{
The inflationary predictions for the quadratic potential model: $n_s$ vs. $r$ (left panel) and $n_s$ vs. $\alpha$ (right panel)
  for various $M_5$ values with $N_0=50$, $60$ and $70$ (from left to right), along with the contours 
  (at the confidence levels of 68\% and 95\%) given by the BICEP2 collaboration (Planck+WP+highL+BICEP2)~\cite{BICEP2}. 
The black triangles are the predictions of the textbook quadratic potential model in the standard cosmology, 
  which are reproduced for $M_5 \gtrsim 1$. 
As $M_5$ is lowered, the inflationary predictions approach the values in Eq.~(\ref{phi2_BClimit}), denoted by the black squares.   
In each line, the turning point appears for $V(\phi_0)/\rho_0 \simeq 1$. 
}
\label{fig:phi2}
\end{figure}

In the brane-world cosmology, these inflationary predictions in the standard cosmology 
 are altered due to the modified Friedmann equation. 
In the limit, $V/\rho_0 \gg 1$,  the Hubble parameter is simplified as $H^2 \simeq  V^2/(3 \rho_0)$, 
 and we can easily find 
\begin{eqnarray}
n_s=1-\frac{5}{2 N_0+1},  \; \; 
r=\frac{24}{2 N_0+1}, \; \; 
\alpha= - \frac{10}{(2 N _0+1)^2}. 
\label{phi2_BClimit}
\end{eqnarray}
The initial ($\phi_0$) and the final ($\phi_e$) inflaton VEVs are found to be 
\begin{eqnarray}
 \phi_0^4 =96 \frac{M_5^6}{m^2} (2 N_0+1), \; \;  \phi_e^4 =96 \frac{M_5^6}{m^2} . 
 \label{phi2_int}
\end{eqnarray}
For a common $N_0$ value, the spectral index is reduced, 
 while $r$ and $|\alpha |$ are found to be larger than those predicted in the standard cosmology. 
In the brane-world cosmology, once $\rho_0$ is fixed, equivalently, $M_5$ is fixed through Eq.~(\ref{rho_0}), 
 the inflaton mass is determined by the constraint ${\cal P}_{\cal S}(k_0)=2.215 \times 10^{-9}$. 
For the limit $V/\rho_0 \gg 1$, we find~\cite{inflation_BC}
\begin{eqnarray}
 \frac{m}{M_5} \simeq 1.26 \times 10^{-4} \left(  \frac{60.5}{N_0+1/2}\right)^{5/6} . 
 \label{m/M5}
\end{eqnarray} 
Note that the model-independent BBN bound on $M_5 > 8.8$ TeV leads to $m > 1.1$ GeV for $N_0=60$.  
Therefore, unlike inflationary scenario in the standard cosmology, the inflaton can be very light, 
 providing the inflationary predictions being compatible with the current observation (see Fig.~\ref{fig:phi2}). 
This analysis is valid for $V(\phi_0)/\rho_0 \gg 1$, in other words, $M_5 \ll 0.01$ 
  by using Eqs.~(\ref{rho_0}), (\ref{phi2_int}), and (\ref{m/M5}).

\begin{figure}[htbp]
\begin{center}
\includegraphics[width=0.45\textwidth,angle=0,scale=1.05]{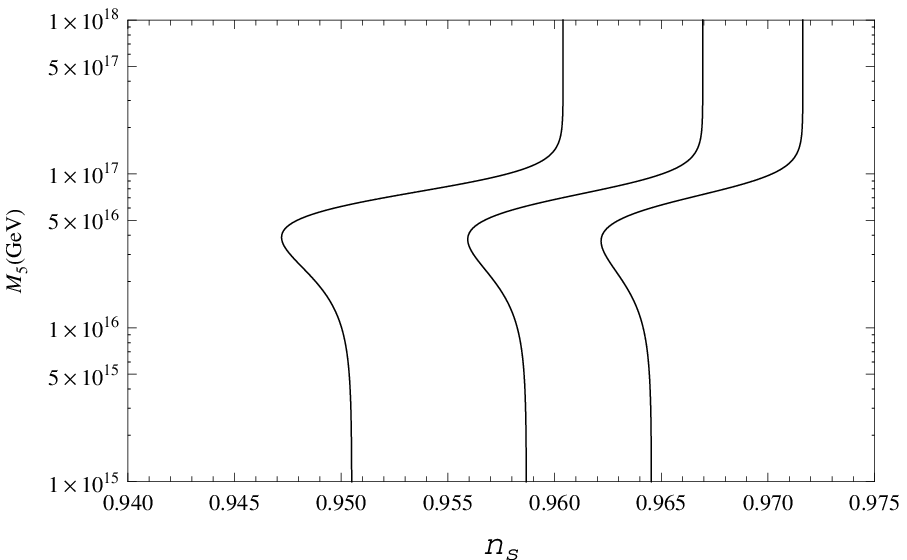} \hspace{0.5cm}
\includegraphics[width=0.45\textwidth,angle=0,scale=1.05]{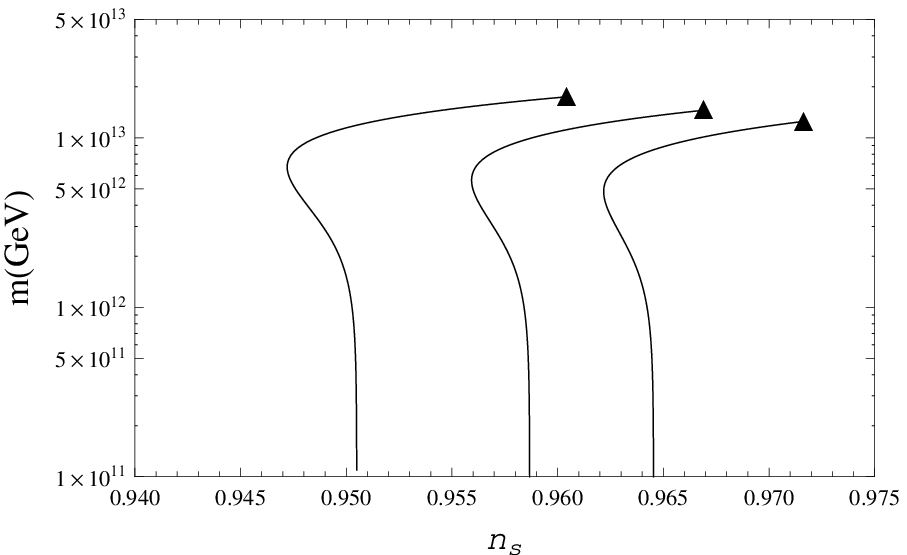}
\end{center}
\caption{
Relations between $n_s$ and $M_5$ (left panel) and between  $n_s$ and $m$ (right-panel), 
  for $N_0=50$, $60$ and $70$ from left to right. 
The black triangles denote the predictions in the standard cosmology.
}
\label{fig:phi2_mass}
\end{figure}

\begin{table}[ht]
\begin{center}
\begin{tabular}{|c|ccccccc|}
\hline \hline
 $M_5 ({\rm GeV})$ & $m ({\rm GeV})$ & $V(\phi_0)^{1/4}$ (GeV)& $\phi_0$ & $\phi_e$ & 
 $n_s$  & $r$ & $-\alpha\ (10^{-4})$ \\
\hline \hline
$ \infty $ & $ 1.46 \times 10^{13} $ & $1.97 \times 10^{16}$ & 
$15.6$ & $1.41 $ & $0.967 $ & $0.132$ & $ 5.46 $  \\ 
\hline
$ 1.02 \times 10^{17}$ & $ 1.38 \times 10^{13} $ & $1.91 \times 10^{16}$ & 
$15.3$ & $1.41 $ & $0.965 $ & $0.160$ & $ 5.50 $  \\
\hline
$ 7.72 \times 10^{16}$ & $ 1.20 \times 10^{13} $ & $1.75 \times 10^{16}$ & 
$14.8$ & $1.41 $ & $0.962 $ & $0.189$ & $ 5.58 $  \\
\hline
$ 5.87 \times 10^{16}$ & $ 9.41 \times 10^{13} $ & $1.49 \times 10^{16}$ & 
$13.7$ & $1.41 $ & $0.958 $ & $0.212$ & $ 6.47 $  \\
\hline
$ 3.79 \times 10^{16}$ & $ 5.69 \times 10^{13} $ & $1.04 \times 10^{16}$ & 
$11.1$ & $1.41 $ & $0.956 $ & $0.219$ & $ 7.28 $  \\
\hline
$ 2.29 \times 10^{16}$ & $ 3.14 \times 10^{12} $ & $6.47 \times 10^{15}$ & 
$7.77$ & $1.40 $ & $0.957 $ & $0.210$ & $ 7.24 $  \\
\hline
$ 2.19 \times 10^{15}$ & $ 2.77 \times 10^{11} $ & $6.30 \times 10^{14}$ & 
$0.834$ & $0.250 $ & $0.959 $ & $0.198$ & $ 6.84 $  \\
\hline
$ 1.23 \times 10^{15}$ & $ 1.55 \times 10^{11} $ & $3.54 \times 10^{14}$ & 
$0.469$ & $0.141 $ & $0.959 $ & $0.198$ & $ 6.83 $  \\
\hline
$ 5.00 \times 10^{14}$ & $ 6.30 \times 10^{10} $ & $1.44 \times 10^{14}$ & 
$0.190$ & $0.0573 $ & $0.959 $ & $0.198$ & $ 6.83 $  \\
\hline
\end{tabular}
\end{center}
\caption{ 
The values of parameters for the potential $V=(1/2) m^2 \phi^2$ for $N_0=60$, 
 in the Planck unit ($M_P=1$) unless otherwise stated. 
} 
\label{table:phi2}
\end{table}

We calculate the inflationary predictions for various values of $M_5$ with fixed e-folding numbers, 
  and show the results in Fig.~\ref{fig:phi2}. 
In the left panel, the inflationary predictions for $N_0=50$, $60$, $70$ from left to right are shown, 
 along with the contours (at the confidence levels of 68\% and 95\%) given by the BICEP2 collaboration 
 (Planck+WP+highL+BICEP2)~\cite{BICEP2}. 
The black triangles represent the predictions of the quadratic potential model in the standard cosmology, 
  while the black squares are the predictions in the limit of $V/\rho_0 \gg 1. $
For various values of $M_5$, the inflationary predictions stay inside of the contour of the BICEP2 result 
 at 95\% confidence level.  
The results for the running of the spectral index ($n_s$ vs. $\alpha$) is shown in the right panel, 
  for $N_0=50$, $60$, $70$ from left to right. 
We also show our results for the 5-dimensional Planck mass ($n_s$ vs. $M_5$) 
  and the inflaton mass ($n_s$ vs. $m$) in Fig.~\ref{fig:phi2_mass}. 
In each solid line, the turning point appears for $V(\phi_0)/\rho_0 \simeq 1$. 
For $N_0=60$, numerical values for selected $M_5$ values are listed in Table~\ref{table:phi2}, for readers convenience.

Next we analyze the textbook quartic potential model, 
\begin{eqnarray}
V =\frac{\lambda}{4!} \phi^4. 
\end{eqnarray}
In the standard cosmology, we find the following inflationary predictions:
\begin{eqnarray}
n_s=1-\frac{6}{2 N_0+3},  \; \; 
r=\frac{32}{2 N_0+3}, \; \; 
\alpha= - \frac{12}{(2 N_0+3)^2}. 
\end{eqnarray}
The quartic coupling ($\lambda$) is determined by the power spectrum measured by the Planck satellite experiment, 
 ${\cal P}_{\cal S}(k_0)=2.215 \times 10^{-9}$ at the pivot scale $k_0=0.05$ Mpc$^{-1}$,  as 
\begin{eqnarray}
  \lambda = 8.46 \times 10^{-13}  \left(  \frac{123}{2 N_0+3}\right)^3. 
\end{eqnarray}

When the limit $V/\rho_0 \gg 1$ is satisfied during the inflation, we find in the brane-world cosmology 
\begin{eqnarray}
n_s=1-\frac{9}{3 N_0+2},  \; \; 
r=\frac{48}{3 N_0+ 2}, \; \; 
\alpha= - \frac{27}{(3 N _0+2)^2}. 
\label{phi4_BClimit}
\end{eqnarray}
The predicted values are very close to those in the standard cosmology for $N_0 \gg 1$. 
Using ${\cal P}_{\cal S}(k_0)=2.215 \times  10^{-9}$, we find 
\begin{eqnarray}
 \lambda = 3.26 \times 10^{-14} \left(  \frac{182}{3 N_0+ 2}\right)^{3} ,  
\end{eqnarray} 
which is independent of $M_5$.

\begin{figure}[htbp]
\begin{center}
\includegraphics[width=0.45\textwidth,angle=0,scale=1]{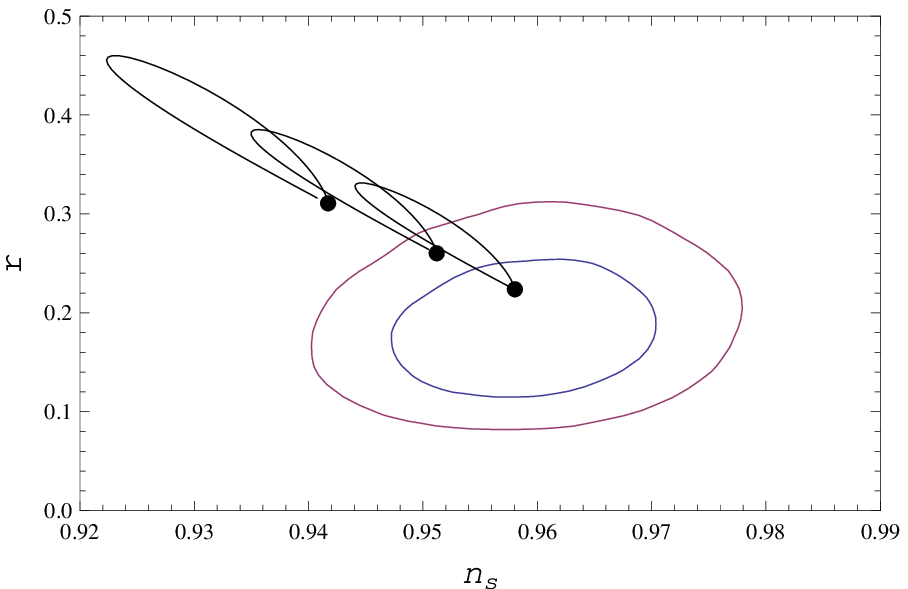} \hspace{0.5cm}
\includegraphics[width=0.45\textwidth,angle=0,scale=1.05]{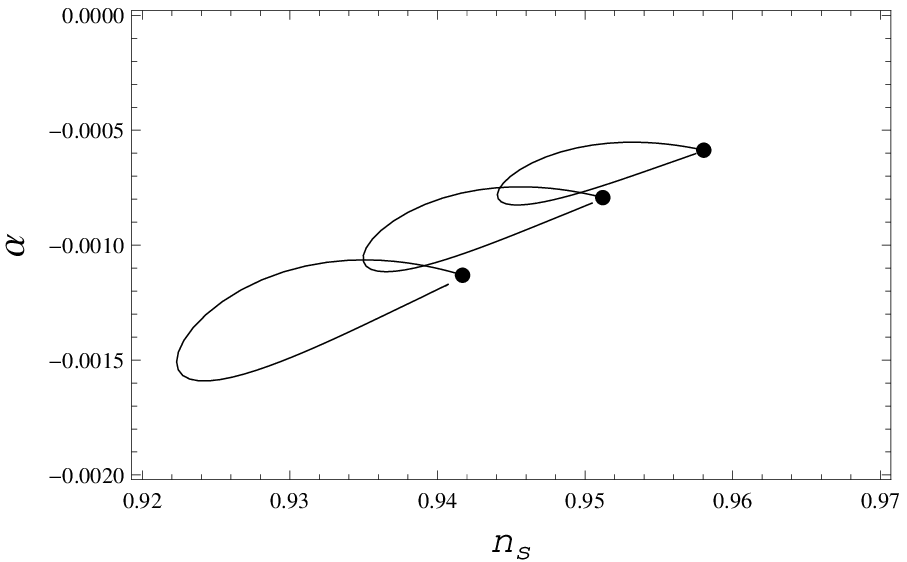}
\end{center}
\caption{
The inflationary predictions for the quartic potential model: $n_s$ vs. $r$ (left panel) and $n_s$ vs. $\alpha$ (right panel)
  with various $M_5$ values for $N_0=50$, $60$ and $70$ (from left to right), along with the contours 
  (at the confidence levels of 68\% and 95\%) given by the BICEP2 collaboration (Planck+WP+highL+BICEP2). 
The black points are the predictions of the textbook quartic potential model in the standard cosmology, 
  which are reproduced for $M_5 \gtrsim 1$. 
As $M_5$ is lowered, the inflationary predictions approach the values in Eq.~(\ref{phi4_BClimit}) 
 which are in fact very close to those in the standard cosmology.  
In each line, the turning point appears for $V(\phi_0)/\rho_0 \simeq 1$. 
}
\label{fig:phi4}
\end{figure}

\begin{figure}[htbp]
\begin{center}
\includegraphics[width=0.45\textwidth,angle=0,scale=1.05]{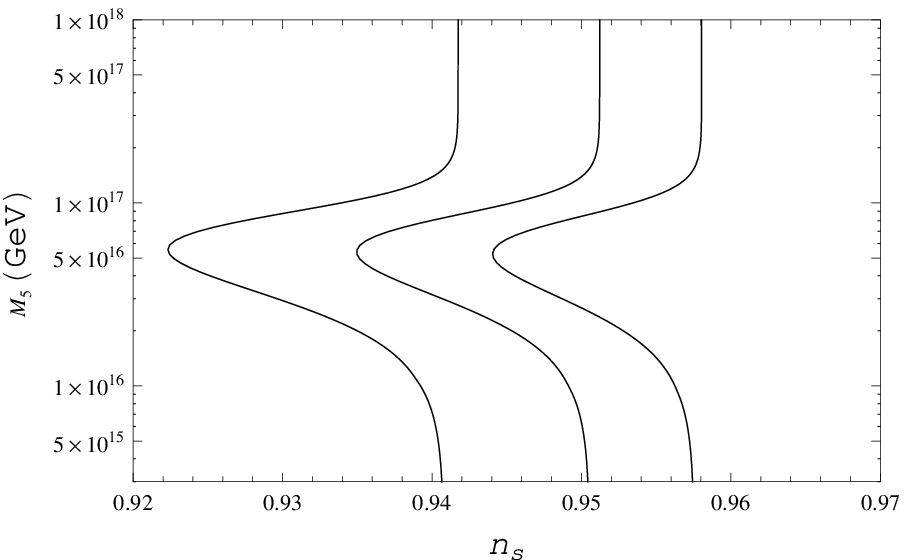} \hspace{0.5cm}
\includegraphics[width=0.45\textwidth,angle=0,scale=1.05]{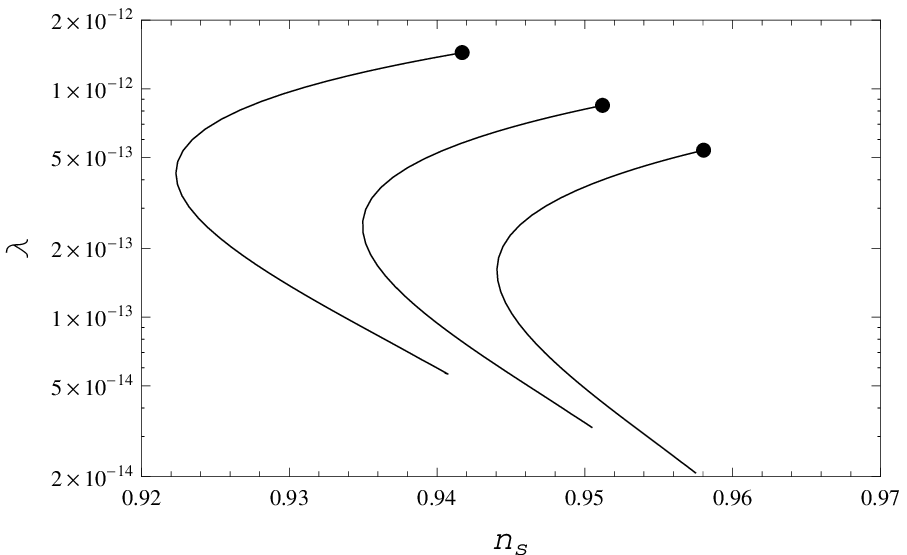}
\end{center}
\caption{
Relations between $n_s$ and $M_5$ (left panel) and between  $n_s$ and $\lambda$ (right-panel), 
  for $N_0=50$, $60$ and $70$ from left to right. 
The black points denote the values obtained in the standard cosmology.
}
\label{fig:phi4_mass}
\end{figure}

We calculate the inflationary predictions for various values of $M_5$ with fixed e-folding numbers. 
Our results are shown in Fig.~\ref{fig:phi4}. 
In the left panel, the inflationary predictions for $N_0=50$, $60$, $70$ from left to right are shown, 
 along with the contours (at the confidence levels of 68\% and 95\%) given by the BICEP2 collaboration 
 (Planck+WP+highL+BICEP2).  
The results for the running of the spectral index ($n_s$ vs. $\alpha$) is shown in the right panel, 
 for $N_0=$50, 60, 70 from left to right. 
 The black points represent the predictions in the standard cosmology.  
In Fig.~\ref{fig:phi4}, the inflationary predictions are moving anti-clockwise along the contours as $M_5$ is lowered. 
The turning point on each contour appears for $V(\phi_0)/\rho_0 \simeq 1$, and as $M_5$ is further lowered, 
 the inflationary predictions go back closer to those in the standard cosmology. 
The brane-world cosmological effect cannot improve the fit of the BICEP2 result.  
We also show corresponding results for the 5-dimensional Planck mass ($n_s$ vs. $M_5$) 
  and the quartic coupling ($n_s$ vs. $\lambda$) in Fig.~\ref{fig:phi4_mass}. 
For $N_0=70$, numerical values for selected $M_5$ values are listed in Table~\ref{table:phi4}.

\begin{table}[ht]
\begin{center}
\begin{tabular}{|c|ccccccc|}
\hline \hline
 $M_5 ({\rm GeV})$ & $ \lambda $ & $V(\phi_0)^{1/4}$ (GeV)& $\phi_0$ & $\phi_e$ & 
 $n_s$  & $r$ & $-\alpha\ (10^{-4})$ \\
\hline \hline
$ \infty $ & $ 5.38 \times 10^{-13} $ & $2.25 \times 10^{16}$ & 
$23.9$ & $3.46 $ & $0.958 $ & $0.224$ & $ 5.87 $  \\ 
\hline
$1.09 \times 10^{17}$ & $4.75 \times 10^{-13}$ & $2.16 \times 10^{16}$ &
$23.7$ & $3.46$ & $0.955$ & $0.272 $ & $5.56$ \\
\hline
$9.07 \times 10^{16} $ & $4.06 \times 10^{-13} $ & $2.05 \times 10^{16}$ & 
$23.3$ & $3.46$ & $0.952$& $0.299$ & $5.57$ \\
\hline
$5.31\times 10^{16} $ &$1.62\times 10^{-13}$ &$1.45 \times 10^{16} $ &
$20.8$ & $3.46$ & $0.944$ & $0.328$ & $7.79$ \\
\hline
$3.39 \times 10^{16}$ & $7.00 \times 10^{-14}$ & $9.69\times 10^{15}$ & 
$17.1$ & $3.46$ & $0.947$ & $0.295$ & $8.11$ \\
\hline 
$2.39\times 10^{16} $ & $4.24\times 10^{-14}$ & $6.88 \times 10^{15}$ & $13.8$ & $3.43$ & $0.951$ & $0.267$ & $7.42$ \\
\hline
$1.59\times 10^{16}$ & $2.93 \times 10^{-14}$ &$4.60 \times 10^{15} $ & 
$10.1$ & $3.24$ & $0.954$ & $0.246$ & $6.71$\\
\hline
$2.20 \times 10^{15}$ & $2.08\times 10^{-14}$ & $6.39 \times 10^{14}$ & 
$1.53$ & $0.698$ &$0.957$ & $0.227$  & $6.02$\\
\hline
\end{tabular}
\end{center}
\caption{ 
The values of parameters for the potential $V=\lambda \phi^4/4!$ for the e-folding number $N_0=70$, 
 in the Planck unit ($M_P=1$) unless otherwise stated. 
} 
\label{table:phi4}
\end{table}

\subsection{Higgs potential model}
Next we consider an inflationary scenario based on the Higgs potential of the form~\cite{Higgs_potential_model}
\begin{eqnarray}
V= \frac{\lambda}{8} \left( \phi^2 -v^2 \right)^2, 
\end{eqnarray}
where $\lambda$ is a real, positive coupling constant, $v$ is a VEV of the inflaton $\phi$. 
For simplicity, we assume the inflaton is a real scalar in this paper, but this model is easily extended to the Higgs model 
  in which the inflaton field breaks a gauge symmetry by its VEV. 
See, for example, Refs.~\cite{hp_corrections, BL_inflation} for recent discussion, 
  where quantum corrections of the Higgs potential are also taken into account.

For analysis of this inflationary scenario, we can consider two cases for the initial inflaton VEVs:
  (i) $\phi_0 < v$ and (ii) $\phi_0 > v$. 
In the case (i), the inflationary prediction for the tensor-to-scalar ratio is found to be small for $v < 1$, since the potential energy 
 during inflation never exceeds $\lambda v^4/8$. 
This Higgs potential model reduces to the textbook quadratic potential model in the limit $v \gg 1$.  
To see this, we rewrite the potential in terms of $\phi=\chi +v$ with an inflaton $\chi$ around the potential minimum at $\phi=v$, 
\begin{eqnarray}
V= \frac{\lambda}{8} \left( 4 v^2 \chi^2 +4 v \chi^3 +\chi^4 \right). 
\end{eqnarray}
It is clear that if a condition, $\chi_0/v \ll 1$, for an initial value of inflaton ($\chi_0$) is satisfied, 
  the inflaton potential is dominated by the quadratic term. 
Recall that $\chi_0 > 1$ in the textbook quadratic potential model and hence $v \gg 1$ is necessary to satisfy the condition. 
In the case (ii), it is clear that the model reduces to the textbook quartic potential model for $ v \ll 1$. 
For the limit $v \gg 1$, we apply the same discussion in the case (i), so that the model reduces to the textbook quadratic potential model. 
Therefore, the inflationary predictions of the model in the case (ii) interpolate the inflationary predictions 
  of the textbook quadratic and quartic potential models by varying the inflation VEV from $v=0$ to $v \gg 1$.

We now consider the brane-world cosmological effects on the Higgs potential model. 
As we have observed in the previous subsection, the inflationary predictions of the textbook models are dramatically altered 
  in the brane-world cosmology, and the quadratic potential model nicely fits the BICEP2 result, while the fitting by the quartic 
  potential model is worse than the fitting in the standard cosmology. 
Thus, in the following, we concentrate on the case (i) with $\phi_0 < v$. 
Even in the brane-world cosmology, the above discussion for (ii) is applicable, namely, 
 the Higgs potential model reduces to  the textbook models for the very small or large VEV limit 
 and  the inflationary predictions for (ii) interpolate those in the two limiting cases. 
Thus, once we have obtained the inflationary predictions for the case (i),  we can imaginary interpolate them 
  to the results for the quartic potential model presented in the previous subsection.

\begin{figure}[htbp]
\begin{center}
\includegraphics[width=0.45\textwidth,angle=0,scale=1.0]{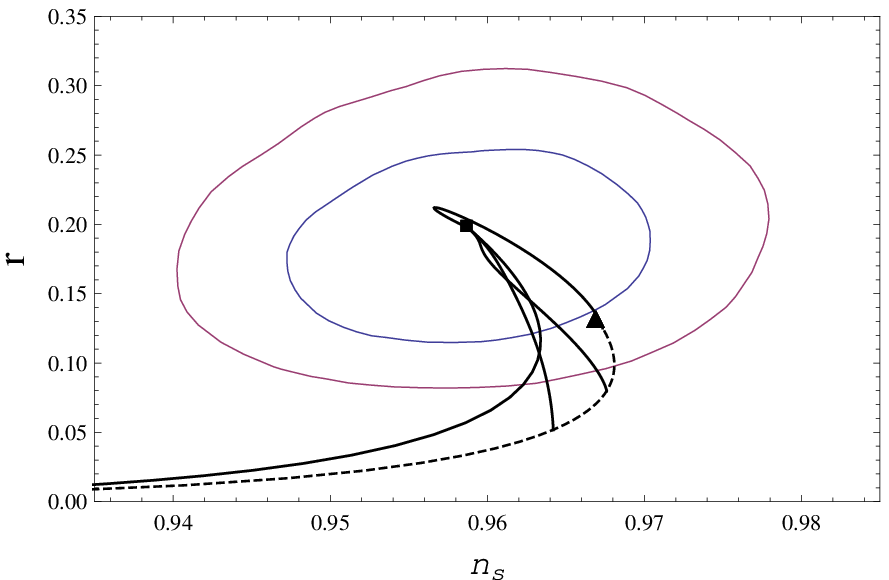} \hspace{0.5cm}
\includegraphics[width=0.45\textwidth,angle=0,scale=1.05]{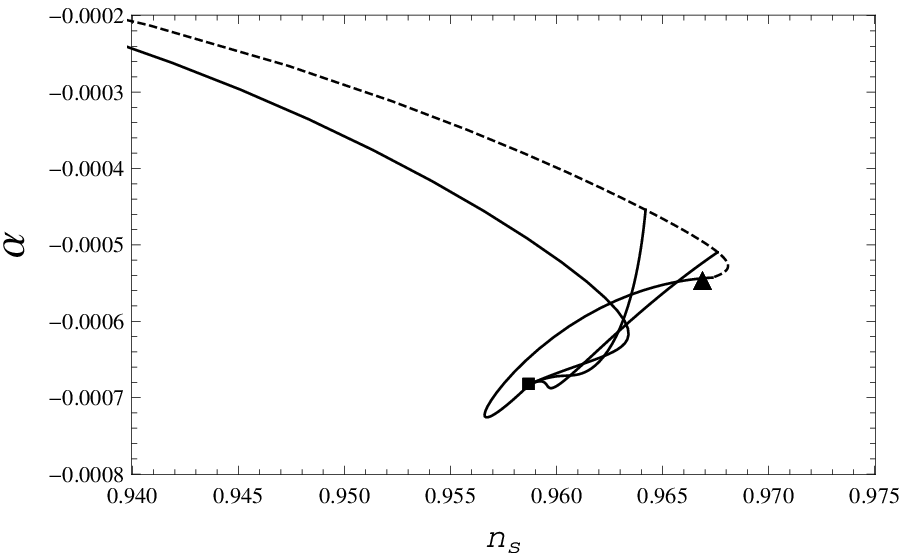}
\end{center}
\caption{
The inflationary predictions for the Higgs potential model: $n_s$ vs. $r$ (left panel) and $n_s$ vs. $\alpha$ (right panel)
  for various $M_5$ values with fixed $v=10$, $20$, $30$ and $200$ from left to right, along with the contours 
  (at the confidence levels of 68\% and 95\%) given by the BICEP2 collaboration (Planck+WP+highL+BICEP2). 
Here we have fixed the number of e-foldings $N_0=60$. 
The dashed line denotes the inflationary predictions for various values of $v$ from $10$ to $200$ in the standard cosmology.
As $v$ is raised, the predicted values of $n_s$ and $r$  approach those of the quadratic potential model along the dashed line 
  (the position marked by the the black triangle).
For $M_5 \gtrsim 1$, the brane-world cosmological effects are negligible, and the predicted values of $n_s$ and $r$ 
  lie on the dashed line. 
As $M_5$ is lowered, the inflationary predictions are deviating from the values on the dashed line, 
 and all solid lines are converging to the point marked by the black squares, which are the predictions 
 of the quadratic potential model for $M_5 \ll 1$. 
}
\label{fig:hp}
\end{figure}

\begin{figure}[htbp]
\begin{center}
\includegraphics[width=0.45\textwidth,angle=0,scale=1.02]{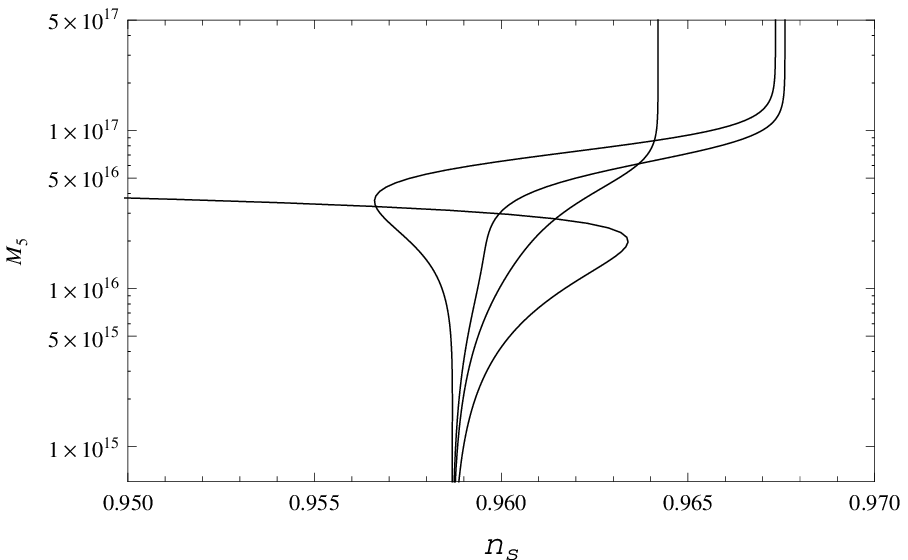} \hspace{0.5cm}
\includegraphics[width=0.45\textwidth,angle=0,scale=1.02]{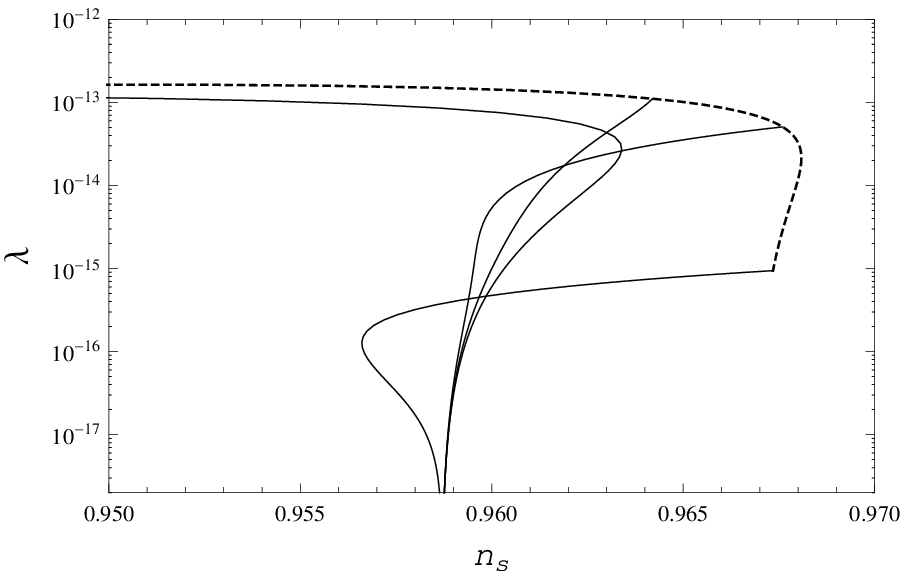}
\end{center}
\caption{
Relations between $n_s$ and $M_5$ (left panel) and between $n_s$ and $\lambda$ (right panel), 
  corresponding to Fig.~\ref{fig:hp}. 
}
\label{fig:hp_mass}
\end{figure}

We calculate the inflationary predictions for various values of $v$ and $M_5$, 
 and the results are shown in Fig.~\ref{fig:hp} for $N_0=60$. 
The dashed line denotes the results in the standard cosmology for various values of $v$. 
The black triangles represent the inflationary predictions in the quadratic potential model, 
  and we have confirmed that the inflationary predictions are approaching the triangles along the dashed line, as $v$ is raised. 
The solid lines show the results for various values of $M_5$ with $v=10$, $20$, $30$ and $200$ from left to right. 
For $M_5 \gtrsim 1$, the brane-world effect is negligible and the inflationary predictions lie on the dashed line. 
As $M_5$ is lowered, the inflationary predictions are deviating from those in the standard cosmology 
  and they approach the values obtained by the quadratic potential model in the brane-world cosmology 
  (shown as the black squares). 
We can see that the brane-world cosmological effect enhances the tensor-to-scalar ratio. 
Fig.~\ref{fig:hp_mass} shows corresponding results in  ($n_s, M_5$)-plane (left panel) and ($n_s, \lambda$)-plane (right panel). 
In the right panel, the dashed line denotes the results in the standard cosmology. 
Numerical values for selected M5 values are listed in Table~\ref{table:hp} 
  for $N_0 = 60$ and $v=20$.
  
\begin{table}[ht]
\begin{center}
\begin{tabular}{|c|ccccccc|}
\hline \hline
 $M_5 ({\rm GeV})$ & $ \lambda $ & $V(\phi_0)^{1/4}$ (GeV)& $\phi_0$ & $\phi_e$ & 
 $n_s$  & $r$ & $-\alpha\ (10^{-4})$ \\
\hline \hline
$ \infty $ & $ 1.11 \times 10^{-13} $ & $1.56 \times 10^{16}$ & 
$7.05$ & $18.6$ & $0.964$ & $0.0519$ & $ 4.54 $  \\ 
\hline
$6.99 \times 10^{16}$ & $8.47 \times 10^{-14}$ & $1.44 \times 10^{16}$ &
$7.59$ & $18.6$ & $0.964$ & $0.0806$ & $5.17$ \\
\hline
$5.08 \times 10^{16} $ & $4.97 \times 10^{-14} $ & $1.23 \times 10^{16}$ & 
$8.68$ & $18.6$ & $0.963$& $0.111$ & $5.95$ \\
\hline
$3.70 \times 10^{16}$ & $2.33 \times 10^{-14}$ & $9.70\times 10^{15}$ & 
$10.3$ & $18.6$ & $0.962$ & $0.138$ & $6.46$ \\
\hline 
$2.01\times 10^{16} $ & $4.78\times 10^{-15}$ & $5.60 \times 10^{15}$ & 
$13.5$ & $18.7$ & $0.961$ & $0.169$ & $6.70$ \\
\hline
$1.85\times 10^{14}$ & $2.29\times 10^{-19}$ &$5.30 \times 10^{13} $ & 
$19.9$ & $20.0$ & $0.959$ & $0.198$ & $6.83$\\
\hline
\end{tabular}
\end{center}
\caption{ 
The values of parameters for the Higgs potential for $N_0=60$ and $v=20$, 
 in the Planck unit ($M_P=1$) unless otherwise stated. 
} 
\label{table:hp}
\end{table}

It is worth mentioning that in the brane-world cosmology, we can realize an inflationary scenario even for $v \ll 1$. 
This is a sharp contrast with the standard cosmological case. 
In the standard cosmology, the slow-roll parameters are given by 
\begin{eqnarray}
\epsilon(\phi_0) = \frac{1}{2} \left(\frac{V'(\phi_0)}{V(\phi_0)} \right)^2=8 \frac{\phi_0^2}{(\phi_0^2-v^2)^2}, \; \; 
\eta(\phi_0) = \frac{V''(\phi_0)}{V(\phi_0)} =4 \frac{3 \phi_0^2-v^2}{(\phi_0^2-v^2)^2}. 
\end{eqnarray}
Hence, for $\phi_0 < v$, $v \gg 1$ is necessary to satisfy the slow-roll conditions, 
 $\epsilon(\phi_0) \ll 1$ and $|\eta(\phi_0)| \ll 1$, simultaneously. 
On the other hand, in the limit of $V/\rho_0 \gg 1$ in the brane-world cosmology, we have 
\begin{eqnarray}
\epsilon(\phi_0) = 12  \frac{M_5^6}{V(\phi_0)} \left(\frac{V'(\phi_0)}{V(\phi_0)} \right)^2, \; \; 
\eta(\phi_0) = 12  \frac{M_5^6}{V(\phi_0)} \left(\frac{V''(\phi_0)}{V(\phi_0)} \right). 
\end{eqnarray}
The slow-roll conditions can be satisfied even for $v \ll 1$ if $M_5 \ll v$.

When the Higgs potential model reduces to the quadratic potential model and $V/\rho_0 \ll 1$, 
 we find from Eq.~(\ref{m/M5}) 
\begin{eqnarray}
 \lambda \simeq 10^{-8} \left(  \frac{M_5}{v}\right)^2. 
\end{eqnarray}
The condition $\chi_0/v \ll 1$ leads to $M_5 \ll 10^{-3} v$ (see Eqs.~(\ref{phi2_int}) and (\ref{m/M5})). 
For $M_5 \sim 10^6$ GeV,  the inflaton has mass of order of the electroweak scale. 
In this case, we may fix the other model parameters as 
$v =10^{10}$ GeV and $\lambda =10^{-16}$ so as to satisfy the above theoretical consistency conditions. 
Because of the brane-world cosmological effect, this inflationary scenario can be consistent 
  with the BICEP2 and Planck results with mass parameters far below the 4-dimensional Planck mass. 
In the next section, we will discuss the reheating process after inflation and implications of such a light inflaton
 to phenomenology at the electroweak scale.

\subsection{Coleman-Weinberg potential}
Finally, we discuss an inflationary scenario based on a potential with a radiative symmetry breaking~\cite{Shafi_Vilenkin} 
  via the Coleman-Weinberg mechanism~\cite{Coleman_Weinberg}. 
We express the Coleman-Weinberg potential of the form, 
\begin{eqnarray}
 V= \lambda \phi^4 \left[ \ln \left( \frac{\phi}{v}\right)-\frac{1}{4}\right]+\frac{\lambda v^4}{4}, 
\end{eqnarray}
where $\lambda$ is a coupling constant, and $ v$ is the inflaton VEV. 
This potential has a minimum at $\phi=v$ with a vanishing cosmological constant. 
The inflationary predictions of the Coleman-Weinberg potential model 
 in the brane-world cosmology have recently been analyzed in ~\cite{CW_BC}. 
However, the modification of the power spectrum of tensor perturbation in Eq.~(\ref{PT}), 
 namely, the function $F$ was not taken into account in the analysis. 
In the following we will correct the results in ~\cite{CW_BC} by taking the function $F$ into account.  
We will see an enhancement of the tensor-to-scalar ratio due to the function $F$.

Analysis for the Coleman-Weinberg potential model is analogous to the one of the Higgs potential model 
   presented in the previous subsection. 
As same in the Higgs potential model, we can consider two cases, (i) $\phi_0 < v$ and (ii) $\phi_0 >v$, for the initial VEV of the inflaton. 
In the same reason as in our discussion about the Higgs potential model, we only consider the case (i) 
  also for the Coleman-Weinberg potential model. 
Since the inflationary predictions in the case (ii) interpolate the predictions of the quartic potential model 
  to those of the quadratic potential model as $v$ is raised (with a fixed $M_5$), 
  one can easily imagine the results in the case (ii).

\begin{figure}[htbp]
\begin{center}
\includegraphics[width=0.45\textwidth,angle=0,scale=1.0]{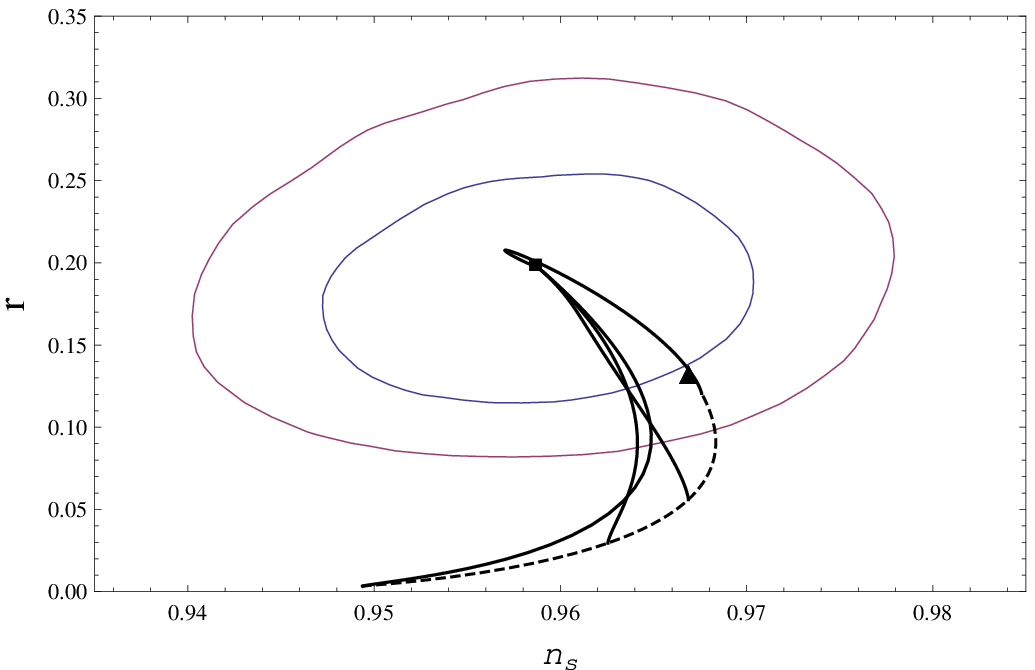} \hspace{0.5cm}
\includegraphics[width=0.45\textwidth,angle=0,scale=1.05]{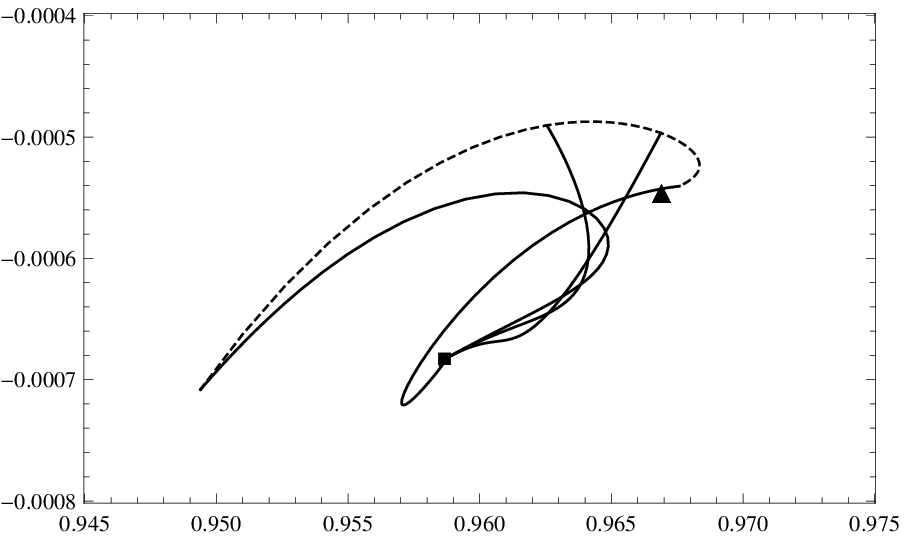}
\end{center}
\caption{
Same as Fig.~\ref{fig:hp} but for the Coleman-Weinberg potential model. 
}
\label{fig:CW}
\end{figure}

\begin{figure}[htbp]
\begin{center}
\includegraphics[width=0.45\textwidth,angle=0,scale=1.02]{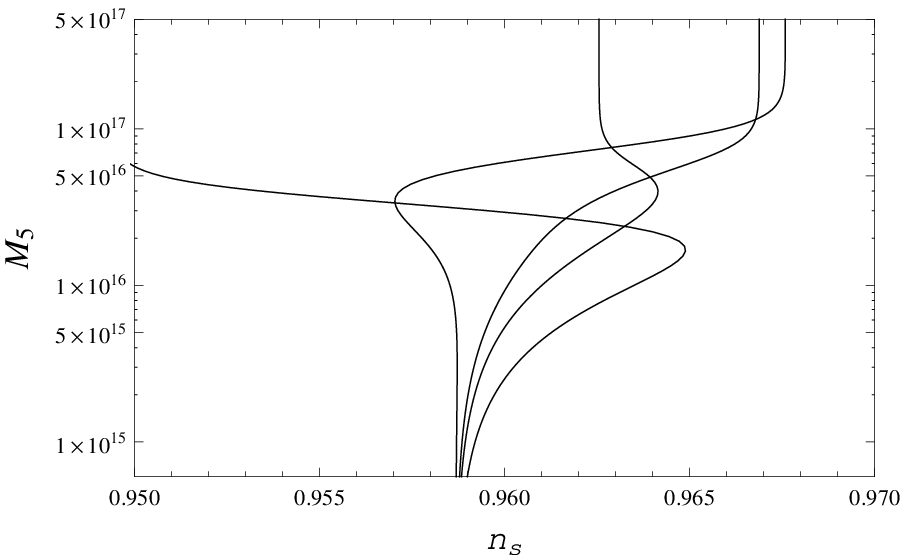} \hspace{0.5cm}
\includegraphics[width=0.45\textwidth,angle=0,scale=1.02]{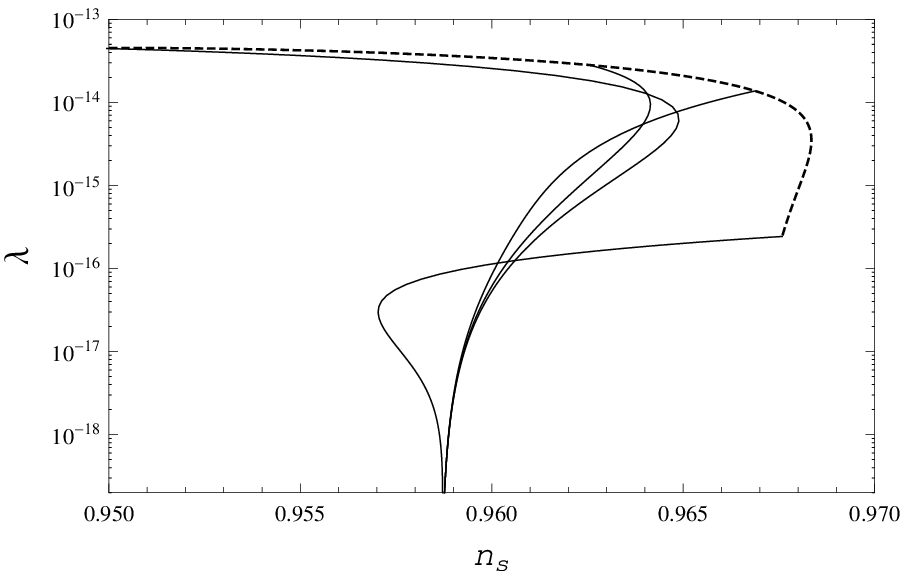}
\end{center}
\caption{
Same as Fig.~\ref{fig:hp_mass} but for the Coleman-Weinberg potential model. 
}
\label{fig:CW_mass}
\end{figure}

We show in Fig.~\ref{fig:CW} the inflationary predictions of the Colman-Weinberg potential model for various values 
 of $v$ and $M_5$ with $N_0=60$.  
The dashed line denotes the results in the standard cosmology for various values of $v$. 
The black triangles denote the results in the quadratic potential model, and we have confirmed that 
  the dashed line approaches the triangles as $v$ is increasing.  
The solid lines show the results for various values of $M_5$ with $v=10$, $20$, $30$ and $200$ from left to right. 
For $M_5 \gtrsim 1$, the brane-world effect is negligible and the inflationary predictions lie on the dashed line. 
As $M_5$ is lowered, the inflationary predictions are deviating from those in the standard cosmology 
   to approach the values obtained by the quadratic potential model in the brane-world cosmology 
  (shown as the black squares). 
Like in the Higgs potential model, for $M_5 \ll v$ the inflationary predictions of the Coleman-Weinberg potential model approach 
 those of the quadratic potential model in the brane-world cosmology. 
Fig.~\ref{fig:CW_mass} shows corresponding results in  ($n_s, M_5$)-plane (left panel) and ($n_s, \lambda$)-plane (right panel). 
In the right panel, the dashed line denotes the results in the standard cosmology. 
For $N_0 = 60$ and $v=20$, numerical values for selected $M_5$ values are listed in Table~\ref{table:CW}.

\begin{table}[ht]
\begin{center}
\begin{tabular}{|c|ccccccc|}
\hline \hline
 $M_5 ({\rm GeV})$ & $ \lambda $ & $V(\phi_0)^{1/4}$ (GeV)& $\phi_0$ & $\phi_e$ & 
 $n_s$  & $r$ & $-\alpha\ (10^{-4})$ \\
\hline \hline
$ \infty $ & $ 2.82 \times 10^{-14} $ & $1.36 \times 10^{16}$ & 
$8.46$ & $18.7$ & $0.963$ & $0.0295$ & $ 4.90 $  \\ 
\hline
$5.38 \times 10^{16}$ & $1.76 \times 10^{-14}$ & $1.19 \times 10^{16}$ &
$9.37$ & $18.7$ & $0.964$ & $0.0617$ & $5.44$ \\
\hline
$3.76 \times 10^{16} $ & $8.19\times 10^{-15} $ & $9.48 \times 10^{15}$ & 
$10.8$ & $18.7$ & $0.964$& $0.0956$ & $5.99$ \\
\hline
$2.49 \times 10^{16}$ & $2.74 \times 10^{-15}$ & $6.72\times 10^{15}$ & 
$12.8$ & $18.7$ & $0.963$ & $0.131$ & $6.34$ \\
\hline 
$1.26\times 10^{16} $ & $4.51\times 10^{-16}$ & $3.53 \times 10^{15}$ & 
$15.8$ & $18.9$ & $0.962$ & $0.166$ & $6.55$ \\
\hline
$1.31\times 10^{14}$ & $2.87\times 10^{-20}$ &$3.75 \times 10^{13} $ & 
$19.9$ & $20.0$ & $0.959$ & $0.198$ & $6.83$\\
\hline
\end{tabular}
\end{center}
\caption{ 
The values of parameters for the Coleman-Weinberg potential for $N_0=60$ and $v=20$, 
 in the Planck unit ($M_P=1$) unless otherwise stated. 
} 
\label{table:CW}
\end{table}

\section{Post inflationary scenario, a light inflaton and phenomenology at the electroweak scale}
For completion of our scenario, we discuss, in this section, post inflationary scenario, namely, 
  reheating process after inflation in the brane-world cosmology.  
In our discussion about reheating, we require the renormalizability of inflationary models, 
  and thus the inflaton only couples with the Higgs doublet among the Standard Model fields. 
In this section, we will consider two example models, the Higgs potential model and the quadratic potential model.

For the Higgs potential model, we introduce the following scalar potential involving the Higgs doublet and inflaton, 
\begin{eqnarray}
V =\frac{1}{8} \lambda (\phi^\dagger \phi -v^2)^2 + \frac{1}{2} \lambda_h \left( \Phi^\dagger \Phi - \frac{v_{\rm EW}^2}{2}\right)^2 
    + \frac{1}{2} \lambda_{\phi h} (\phi^\dagger \phi -v^2) \left( \Phi^\dagger \Phi- \frac{v^2}{2}\right), 
\end{eqnarray}
where $\phi$ is the inflaton field, $\Phi$ is the Standard Model Higgs doublet, 
 $v_{\rm EW}=246$ GeV is the Higgs VEV in the Standard Model, and $\lambda$s are quartic coupling 
 constants being real and positive.\footnote{ 
Here we have restricted the scalar potential to simplify our discussion. 
If the inflaton field is a complex scalar in some gauge extension of the model, this scalar potential 
 is the most general, renormalizable one with a vanishing cosmological constant. 
However, the inflaton is a real scalar in our simple setup, and one can add, in general, more terms 
  such as $\phi (\Phi^\dagger \Phi)$ and $\phi^3$ to the scalar potential.
}
This potential has a vacuum at $\langle \phi \rangle =v$ and $\langle \Phi_0 \rangle =v_{\rm EW}/\sqrt{2}$, 
  where $\Phi_0$ is an electric charge neutral component in the Higgs doublet. 
We rewrite the potential around the vacuum with $\phi=\chi +v$ and $\Phi_0=(h+v_{\rm EW})/\sqrt{2}$, 
\begin{eqnarray}
  V&=&\frac{1}{2} m_\phi^2 \chi^2 +\frac{1}{2} m_h^2 h^2 + m_{\phi h}^2 \chi h 
 + \frac{m_\phi^2}{2 v} \chi^3 + \frac{m_{\phi h}^2}{2v} \chi^2 h +\frac{m_{\phi h}^2}{2v_{\rm EW}} \chi h^2+\frac{m_h^2}{2 v_{\rm EW}}h^3 
 \nonumber\\
 &+& \frac{1}{8} \left( \frac{m_\phi}{v} \right)^2 \chi^4 + \frac{1}{8} \left(\frac{m_h}{v_{\rm EW}}\right)^2 h^4 
  + \frac{1}{4} \left( \frac{m_{\phi h}^2}{v v_{\rm EW}} \right) \chi^2 h^2, 
\label{potential2}
\end{eqnarray}
where we have introduced three mass parameters given by
 $m_\phi=\sqrt{\lambda}v$, $m_h=\sqrt{\lambda_h} v_{\rm EW}$ and $m_{\phi h}=\sqrt{\lambda_{\phi h}  v_{\rm EW} v}$.

The inflaton mixes with the Standard Model Higgs boson through the mass term $m_{\phi h}^2 \chi h$. 
Here we consider the case in which the mixing is so small that $h$ is almost identical to the Standard Model Higgs boson. 
In this case, the mass matrix of $\phi$ and $h$  is diagonalized by the mass eigenstates $\Phi_1$ and $\Phi_2$ 
 defined as 
\begin{eqnarray}
   h \simeq \Phi_1 + \delta \; \Phi_2, \; \;  \chi \simeq - \delta\;  \Phi_1 + \Phi_2, 
\label{mass_eigenstate}
\end{eqnarray}
with the mass eigenvalues, $m_1 \simeq m_h$ and $m_2 \simeq m_\phi$,  and a small mixing parameter $|\delta| \ll 1$ defined as 
\begin{eqnarray}
 \delta \simeq \frac{m_{\phi h}^2}{m_\phi^2-m_h^2}. 
\end{eqnarray}

Now we consider reheating process through the inflaton decay. 
When the inflaton is much heavier than the Higgs boson ($m_1  \ll m_2$), the inflaton decays to a pair of the Higgs doublets 
 through the interaction
\begin{eqnarray}
 {\cal L}_{\rm int} \supset -\frac{1}{2} \lambda_{\phi h} v \phi (\Phi^\dagger \Phi) \simeq 
 -\frac{1}{2} \delta \frac{m_\phi^2}{v_{\rm EW}} \phi (\Phi^\dagger \Phi) .
\end{eqnarray}
For the process $\phi \to \Phi^\dagger \Phi$,  we have the decay width 
\begin{eqnarray}
\Gamma(\phi \to \Phi^\dagger \Phi) = \frac{1}{8 \pi} \left( \delta  \frac{m_\phi}{v_{\rm EW}}\right)^2 m_\phi. 
\end{eqnarray}
In evaluating reheating temperature, we compare this decay width with the Hubble parameter 
\begin{eqnarray}
 H=\sqrt{\frac{1}{3} \rho \left( 1+ \frac{\rho}{\rho_0} \right)} \simeq \frac{\rho}{6 M_5^3}, 
\end{eqnarray}
where $\rho=(\pi^2/30) g_* T_{\rm RH}^4$ with a total number of degrees of freedom ($g_*\simeq 100$). 
For example, if we take $m_\phi=1$ TeV and $\delta =0.01$, we find $T_{\rm RH} \simeq 4.9 \times 10^{4}$ GeV, 
 by using Eq.~(\ref{m/M5}).

Since in the brane-world cosmology the inflaton mass can be as low as the electroweak scale, 
 let us consider a radical case with $m_\phi < m_h/2$, so that the Standard Model-like Higgs boson 
 can decay to a pair of inflatons.  
Rewriting the scalar potential in Eq.~(\ref{potential2}) by the mass eigenstate of Eq.~(\ref{mass_eigenstate}), 
 we find an interaction term, 
\begin{eqnarray}
 {\cal L}_{\rm int} \supset -\frac{1}{2}\left( \frac{2 m_\phi^2+m_h^2}{v_{\rm EW}} \right)\delta^2 \Phi_1 \Phi_2^2, 
\end{eqnarray}
where $\Phi_1$ is the Standard Model-like Higgs boson, and $\Phi_2$ is the inflaton. 
Thus we have a partial decay width of the Higgs boson to a pair of inflatons as  
\begin{eqnarray}
   \Gamma(\Phi_1 \to \Phi_2 \Phi_2) \simeq \frac{\delta^4}{32 \pi m_h} \left( \frac{2 m_\phi^2+m_h^2}{v_{\rm EW}} \right)^2 
   \sqrt{1-\frac{4 m_\phi^2}{m_h^2}}. 
\end{eqnarray}
When we take as reference values  $m_\phi=50$ GeV and $\delta =0.105$ for a Higgs boson mass $m_h=125$ GeV, 
 we find $\Gamma(\Phi_1 \to \Phi_2 \Phi_2) \simeq 0.01 \Gamma_h$, where $\Gamma_h = 4.07$ MeV is 
 the total Higgs boson decay width~\cite{Gamma_h}.
Once a Higgs boson is produced in collider experiments, it decays to a pair of inflatons with 1\% branching fraction. 
Then, each produced inflaton mainly decays to $b \bar{b}$ through the mixing with the Standard Model Higgs boson 
  (see the next paragraph for discussion about this decay process). 
Note that this process, $\Phi_1 \to \Phi_2 \Phi_2 \to 2 b 2{\bar b}$, is similar to the Higgs boson pair production, 
 followed by the decay of each Higgs boson to $b \bar{b}$. 
Precision measurement of the Higgs boson properties is one of major tasks of future collider experiments such as the ILC. 
Here we refer ILC studies~\cite{ILC_Higgs} on the measurement of the Higgs boson self-coupling 
 through a Higgs pair production process associated with $Z$ boson, 
 $e^+e^- \to Z^* \to Z h^* \to Z h h $, followed by $h \to b {\bar b}$.  
The production cross section of this process is $\sim 0.1$ fb for a collider energy $0.5-1$ TeV. 
On the other hand, in our case, a pair of inflatons can be produced via $e^+e^- \to Z^* \to Z \Phi_1$, followed by 
  the Standard Model-like Higgs boson decay to a pair of inflatons, $\Phi_1 \to \Phi_2 \Phi_2$. 
Since the Higgs boson production associated with $Z$ boson has a cross section$\sim 10$ fb, 
  the inflaton pair production cross section from the 1\% branching fraction of the Higgs boson decay 
  is comparable to the Higgs pair production cross section with the same final states. 
Therefore, precision measurements of the Higgs self-coupling at the ILC can reveal the inflaton in the brane-world cosmology, 
  although it is difficult to distinguish the inflaton from a light singlet scalar added to the Standard Model.

Now we go back to discussion about reheating.   
When the inflaton is lighter than the Higgs boson, its main decay mode is to $b {\bar b}$ through the mixing with the Higgs boson. 
The decay width is calculated to be
\begin{eqnarray}
\Gamma (\Phi_2 \to b {\bar b} )\simeq \frac{3}{8 \pi} \delta^2 \left(  \frac{m_b}{v_{\rm EW}}\right)^2 m_\phi 
  \left(1-\frac{4 m_b^2}{m_\phi^2} \right)^{3/2}, 
\end{eqnarray}
where $m_b=4.2$ GeV is the bottom quark mass. 
Using the condition,  $\Gamma (\Phi_2 \to b {\bar b} )=H(T_{\rm RH})$,  in the limit $\rho/\rho_0 \gg 1$, 
 we find $T_{\rm RH} \simeq 676$ GeV for $m_\phi=50$ GeV, $\delta=0.105$ and $M_5 =4.0 \times 10^5$ GeV (see Eq.~(\ref{m/M5})). 
In this case, the "transition temperature" ($T_t$), at which the evolution of universe transits from the brane-world cosmology 
 to the standard cosmology, is calculated as $T_t \simeq 0.1$ GeV by $\rho(T_t)/\rho_0=1$.
The transition temperature is so low that the brane-world cosmological effects dramatically alter 
  the results obtained in the standard particle cosmology~\cite{DM_BC, LG_BC, gravitino_BC}.

The next example is the quadratic potential model. 
When we introduce general renormalizable terms between the inflaton and Higgs doublet such as 
  $\phi \Phi^\dagger \Phi$, our discussion is analogous to that for the first example. 
Here, let us examine a novel scenario, namely, a unified picture of inflaton and the dark matter particle. 
For $M_5 \sim 10^6$ GeV, the inflaton has mass at the electroweak scale, 
 providing the inflationary predictions being consistent with the current observations. 
When we introduce a $Z_2$ parity and assign odd-parity for the inflaton while even-parity 
  for all of the Standard Model particles, the inflaton with the coupling to the Higgs doublet 
  plays the role of the so-called Higgs portal dark matter~\cite{Higgs_portal_DM}. 
A unified picture of the inflaton and the Higgs portal dark matter has been realized in an inflationary model 
  with non-minimal gravitational coupling~\cite{WIMP_inflaton1, WIMP_inflaton2}. 
Now we consider the unified picture in the brane-world scenario.

Because of the $Z_2$ invariance and the requirement of the renormalizability, 
 a unique interaction term of the inflaton with the Higgs doublet is 
\begin{eqnarray}
 {\cal L}_{\rm int} = g^2 \phi^2 |\Phi|^2 , 
 \label{I-H int}
\end{eqnarray}
where $g$ is a coupling constant.\footnote{
This term induces an inflaton quartic coupling $\sim g^4/(4 \pi)^2$ at the quantum-level. 
We practically adjust an inflaton quartic coupling at the tree-level to cancel the induced 
  coupling to make an effective quartic coupling negligibly small. 
Thus, the inflation is  governed by the quadratic potential.   
}
Through this interaction term, a pair of the inflatons annihilate into the Standard Model particles 
  via the Higgs boson exchange in the $s-$channel and, if kinematically allowed, 
  into a pair of the Higgs bosons.  
For the inflaton/dark matter mass $\sim$TeV (which means $M_5 \sim 10^7$ GeV)  for example, 
  $g^2 \sim 0.1$ reproduces the observed relic abundance of the dark matter particle 
  (see, for example, \cite{KMNO} for analysis of the dark matter relic abundance 
  and the direct dark matter detection experiments, as well as the production of the Higgs portal dark matter particle 
  at the Large Hadron Collider).\footnote{
Precisely speaking, this result in the standard cosmology should be altered in the brane-world cosmology. 
However, for $M_5 \sim 10^7$ GeV, the transition temperature is close to the decoupling temperature 
  of the dark matter particle, and the brane-world cosmological effect is not significant~\cite{DM_BC}. 
}

After inflation, the inflaton oscillates around the potential minimum. 
During this oscillation phase, the inflaton energy density is transmitted to relativistic particles to thermalize the universe. 
Since the inflaton cannot decay due to the $Z_2$-parity, preheating~\cite{preheating} through the interaction 
 in Eq.~(\ref{I-H int}) plays the crucial role. 
We expect an explosive Higgs doublet production through the parametric resonance 
  during the oscillation phase and the decay products of Higgs doublet get  the universe thermalized~\cite{WIMP_inflaton2}. 
Reheating temperature in this case is estimated by 
\begin{eqnarray}
  \Gamma_h = H(T_{\rm RH}) \simeq \frac{\rho(T_{\rm RH})}{6 M_5^3}, 
\end{eqnarray}
where $\Gamma_h=4.07$ MeV is the total Higgs boson decay width. 
We find $T_{\rm RH} \simeq 2.9 \times 10^4$ GeV, which is high enough to get the inflaton in thermal equilibrium, 
   and the standard discussion about thermal relic dark matter particle follows.\footnote{
Our discussion about preheating and subsequent thermalization of the universe here 
 is naive and in fact, these processes are very complicated. 
For detailed analysis, see, for example,  Ref.~\cite{PreheatDetail}.  }

\section{Conclusions}
Motivated by the recent observation of the CMS $B$-mode polarization by the BICEP2 collaboration, 
  we have studied simple inflationary models based on the quadratic, quartic, Higgs and Coleman-Weinberg potentials 
  in the context of the brane-world cosmology. 
For the 5-dimensional Planck mass $M_5 < M_P$, the brane-world cosmological effect 
  alters inflationary predictions from those in the standard cosmology. 
We have found that all simple models except the quartic potential model can fit the results by 
 the BICEP2 and Planck satellite experiments with an enhancement of the tensor-to-scalar ratio 
 in the presence of the 5-dimensional bulk. 
Keeping the inflationary predictions to be consistent with the observations, 
  the inflaton can become lighter as the 5-dimensional Planck mass is lowered. 
Requiring the renormalizability of the inflationary models, the inflaton only couples 
  with the Higgs doublet among the Standard Model fields, and the coupling plays the crucial role 
  in reheating  process after inflation. 
When an inflaton has mass at the electroweak scale,  this light inflaton has some impacts 
  on phenomenology at the electroweak scale. 
We have discussed two scenarios.   
In the first scenario, the Higgs boson decay can have a 1\% branching fraction to a pair of inflatons,
  which can be tested in the future collider experiments.  
The second scenario offers a unified picture of the inflaton and the dark matter particle, 
  where  the light inflaton plays the role of the Higgs portal dark matter.

Finally we comment on a tension between the BICEP2 result of $r \simeq 0.2$  
  and a constraint by the Planck measurement on $r < 0.11$.  
As discussed by the BICEP2 collaboration~\cite{BICEP2}, a way to reconcile the two results 
  is to introduce the running spectral index $\alpha=d n_s/d \ln k \simeq -0.022$. 
The contours given by the BICEP2 collaboration (Planck+WP+highL+BICEP2) 
  shown in Figures in this paper are obtained by setting $\alpha \simeq -0.022$. 
On the other hand, as we have shown in Figures, the simple inflationary models predict  
 $-\alpha={\cal O}(10^{-3})$, which is too small to make the BICEP2 result compatible with 
 the constraint quoted by the Planck measurement.  
Thus, rigorously speaking, inflationary predictions with $r \sim 0.2$ in the simple modes we have examined are not the best fit. 
We may regard the contours given by the BICEP2 collaboration as a reference, and the tensor-to-scalar ratio $r \sim 0.1$ 
   may be more reasonable to discuss the fitting of the BICEP2 and Planck results. 
In this regard, the experimental results favor the Higgs potential and the Coleman-Weinberg potential  
  models with a mild brane-world cosmological effect, $V/\rho_0 \lesssim1$ 
  (see Figs.~\ref{fig:hp} \& \ref{fig:CW} and Tables~\ref{table:hp} \& \ref{table:CW}).

\section*{Acknowledgments}
We would like to thank Andy Okada for his encouragement. 


\end{document}